\documentstyle[12pt]{article}
\begin{document}
{\small
\rightline{hep-th/9612176}
\rightline{IC/96/268}
}
\vskip 1.0 true cm
{\large
\centerline{Self-duality in Generalized Lorentz Superspaces}
\vskip 0.7 true cm
\centerline{C. Devchand$^1$\ \ and\ \  J. Nuyts$^2$}}
\vskip 0.6 true cm
{\small
\centerline{devchand@ictp.trieste.it , nuyts@umh.ac.be}
\vskip 0.2 true cm
\centerline{$^1$ International Centre for Theoretical Physics,
34100 Trieste, Italy}

\centerline{$^2$ Universit\'e de Mons-Hainaut, 
20 Place du Parc, 7000 Mons, Belgium}
}
\vskip 1 true cm
\noindent
{\bf Abstract}
  
\noindent
We extend the notion of self-duality to spaces built from a set
of representations of the Lorentz group with bosonic or fermionic
behaviour, not having the traditional spin-one upper-bound of super
Minkowski space.  The generalized derivative vector fields on such 
superspaces are assumed to form a superalgebra. Introducing
corresponding gauge potentials and hence covariant
derivatives and curvatures, we define generalized self-duality as the
Lorentz covariant vanishing of certain irreducible parts of the
curvatures. 

\section{Introduction}

Self-duality (or antiself-duality) has been an important tool in the
study of gauge theories and is well-known to have led to many
profound and successful results. Some years ago \cite{CDFN},
self-duality was generalized to spaces of higher dimensions by
introducing a completely antisymmetric four-index tensor. Here, we
generalize self-duality in another direction keeping, this time,
Lorentz covariance. 

We introduce, in a Lorentz covariant fashion, generalized superspaces 
with bosonic and fermionic coordinates $\{Y\}$. Corresponding to each
coordinate, we define an operator $X$ (generalizing the derivative) 
which transforms under the Lorentz group in the same way as the
coordinate. These generalized derivatives are assumed to form a
superalgebra and to act linearly on the coordinates. 
Then, by associating a gauge potential $A$ to each of these generalized
derivatives, we define, in a natural fashion, the covariant derivative 
${\cal{D}}=X+A$ and the corresponding curvatures $F$.
Generalized self-dualities are then defined in terms of the Lorentz
covariant conditions of the vanishing of curvature components
corresponding to subsets of specified Lorentz behaviour.

These generalized superspaces are modeled on standard superspace
with coordinates $\{Y^{\alpha \dot\alpha},Y^{\alpha},Y^{\dot\alpha}\}$ 
(see the next section for notation) and super derivative operators 
$\{X_{\alpha \dot\alpha},$ $X_{\alpha}, X_{\dot\alpha}\}$.
They provide, in particular, higher dimensional spaces having manifest
four-dimensional Lorentz covariance.

\section{Multi-index notation}

We use the dotted ($\dot \alpha=1,2$) and undotted ($\alpha=1,2$)
indices familiar for the Lorentz group. These can be lowered (or
raised) using the antisymmetric $\epsilon_{\alpha\beta}$ and
$\epsilon_{\dot\alpha\dot\beta}$ tensors with $\epsilon_{12}=1$ (or
$\epsilon^{\alpha\beta}$ and $\epsilon^{\dot\alpha\dot\beta}$ with
$\epsilon^{12}=-1$). 

We define the multi-indices $[A],[B],[\dot A]$ and $[\dot B]$ as the 
sets of, respectively, $2a,2b,2\dot a$ and $2\dot b$ symmetrized 
indices ($a, b, \dot a$ and $\dot b$ being integers or half-integers),
\begin{eqnarray}
[A]=\alpha_1\alpha_2\ldots\alpha_{2a}\ \ ,&\quad&
[B]=\beta_1 \beta_2\ldots\beta_{2b}\ \  ,
\nonumber \\[5pt]
[\dot{A}]=\dot{\alpha_1}\dot{\alpha_2}
     \ldots\dot\alpha_{2\dot{a}}\ \ ,&\quad&
[\dot B] = \dot{\beta_1}\dot{\beta_2}\ldots \dot\beta_{2\dot b}
\ \  .\label{sets}
\end{eqnarray}

Using the $\epsilon$'s we also define the multi-index tensors
\begin{eqnarray}
\epsilon_{[s]}&=&\epsilon_{\alpha_1\beta_1}
                  \epsilon_{\alpha_2\beta_2}
                  \ldots
                  \epsilon_{\alpha_s\beta_s}  
\nonumber \\
\epsilon^{[\dot t]}&=&
         \epsilon^{\dot\alpha_1\dot\beta_1}
         \epsilon^{\dot\alpha_2\dot\beta_2}
         \ldots
         \epsilon^{\dot\alpha_{\dot t}\dot\beta_{\dot t}} 
\\  \epsilon^{[0]}&=& \epsilon_{[0]}= 1
\ \  .
\nonumber
\label{genep}
\end{eqnarray}

Finally, $S[A]$ denotes the symmetrization operator, i.e. the
sum over all permutations of indices appearing in $[A]$.

\section{Coordinates}

The generalized coordinates $\{Y\}$ are a set of Lorentz tensors
$Y_{[A]}^{[\dot A]}$ transforming as $(a,\dot a)$ representations of
the Lorentz group. These representations are called even (bosonic) if
$2(a+\dot a)$ is even and odd (fermionic) if $2(a+\dot a)$ is odd.
They commute or anticommute according to the obvious rules
\begin{equation}
\left [Y_{[A]}^{[\dot A]} ~,~ Y_{[B]}^{[\dot B]}\right ]_{\eta}=0\ , 
\label{comYY}
\end{equation}
where the graded bracket is defined with
\begin{equation}
\eta=(-1)^{4(a+\dot a)(b+\dot b)+1}\  .
\label{eta} 
\end{equation}

Note that the multiplicity of a given representation can be any
non-negative integer. Traditional superspaces possess only a finite
number of representations, each appearing possibly more than once. 
To simplify the present exposition, we shall not introduce a
multiplicity index. In other words, we assume that each representation
appears only once. The generalization to representations with
multiplicities is essentially obvious and will be discussed in 
\cite{DN}. 

\goodbreak
\section{Generalized derivatives}

Let us associate to the coordinates $\{Y\}$, generalized derivatives 
$\{X\}$ satisfying a Lorentz covariant superalgebra $\cal{A}$.
Using the multi-indices
(for $0\leq s \leq \min(2a,2b)$ and for
$0\leq \dot t \leq \min(2\dot a,2\dot b)$)
\begin{equation}\begin{array}{rll}
[C(s)]&=&
          \alpha_{s+1}
          \alpha_{s+2}
          \ldots
          \alpha_{2a}
          \beta_{s+1}
          \beta_{s+2}
          \ldots
          \beta_{2b}  
\ \  ,  \\[5pt]
\left [\dot C(\dot t)\right ]&=&
               \dot\alpha_{\dot t+1}
               \dot\alpha_{\dot t+2}
               \ldots
               \dot\alpha_{2\dot a}
               \dot\beta_{\dot t+1}
               \dot\beta_{\dot t+2}
               \ldots
               \dot\beta_{2\dot b} 
\ \  ,
\label{Csets}
\end{array}\end{equation}
the Lorentz--allowed supercommutation relations of the $X$'s are 
\begin{eqnarray}
 \left [X_{[A]}^{[\dot{A}]} ~,~ X_{[B]}^{[\dot{B}]}\right ]_{\eta} 
 & = &
     \sum_{s=0}^{\min (2a,2b)}\sum_{\dot t=0}^{\min (2\dot a,2\dot b)}
     t(a,\dot a;b,\dot b;a+b-s,\dot a+\dot b-\dot t)
\\ && \qquad \qquad \qquad \qquad
     S[A]S[\dot A]S[B]S[\dot B]
     \epsilon_{[s]}\epsilon^{[\dot t]}
     X_{[C(s)]}^{[\dot C(\dot t)]}
\label{comXX}\ \  ,\nonumber
\end{eqnarray} 
where the $t(a,\dot a\ ;b,\dot b\ ;c,\dot c)$'s depending on six 
integers or half-integers are the structure constants of the 
superalgebra and $\eta$ is again given by (\ref{eta}). Obviously, 
the allowed values of the structure constants are restricted by 
(skew-)symmetry and by super Jacobi identities. A similar class 
of algebras has been considered previously by Fradkin and Vasiliev
\cite{FV}.

It is natural but not compulsory to identify the generators of
the Lorentz group itself with the six generators in the set $\{X\}$
behaving as $(1,0)\oplus(0,1)$.

\section{Action of the generalized derivatives on the coordinates} 

We take the elements of the algebra $\cal{A}$ as acting on the space
coordinates $Y$ as generalized derivatives. This means that the
supercommutation relations between the differential operators $X$ and
the coordinates $Y$ are taken to be linear in the $Y$'s i.e. the
$X$'s transform the coordinates at most linearly among themselves. 
The $X$'s together with the $Y$'s therefore combine to form an
enlarged superalgebra.  Let us write, using the by now familiar
notation,
\begin{eqnarray}
&&\left [X_{[A]}^{[\dot A]} ~,~ Y_{[B]}^{[\dot B]}\right ]_{\eta} 
\nonumber \\
&& = \sum_{s=0}^{\min (2a,2b)}\sum_{\dot t=0}^{\min (2\dot a,2\dot b)}
  u(a,\dot a\ ;b,\dot b\ ;a+b-s,\dot a+\dot b-\dot t)
\nonumber \\
&& \qquad\qquad \qquad \qquad \quad
S[A]S[\dot A]S[B]S[\dot B]\epsilon_{[s]}\epsilon^{[\dot t]}
     Y_{[C(s)]}^{[\dot C(\dot t)]}    
\nonumber \\
&& \quad      +\ c(a,\dot a) \delta_{ab} \delta^{\dot a\dot b}
   S[A]S[\dot A]\epsilon_{[2a]}\epsilon^{[2\dot a]}
\ \ \ .
\label{comXY}
\end{eqnarray} 
The $u$'s are new structure constants. The central $c$-terms can
be either zero or, if they are non-zero and if there are no
multiplicities, can be set to 1 by a suitable renormalization of the
operators. The allowed values of the $u$'s and the $c$'s are
restricted by the super-Jacobi identities for the combined
superalgebra of the $X$'s and the $Y$'s. 

\section{Gauge fields}

To every operator $X$ we associate a generalized gauge potential $A$
having the same Lorentz transformation properties and depending on
the variables $Y$. These gauge potentials take values in the algebra of
some matrix group whose $N$ generators are denoted 
$\lambda_k,\   k=1,\ldots,N$, in some representation defined and 
normalized so that
\begin{eqnarray}
\left [\lambda_k,\lambda_l\right ]&=&f_{kl}^{{\phantom{kl}}m}\lambda_m  
\nonumber \ \  ,\\
{\rm{tr}}(\lambda_k\lambda_l)&=&\delta_{kl}\  \  ,
\label{Liegenerators}\end{eqnarray}
where $f_{kl}^{{\phantom{kl}}m}$ are the structure constants.
We have
\begin{eqnarray}
A_{[A]}^{[\dot A]}&=&\sum_{k=1}^{N} A_{[A],k}^{[\dot A]}\lambda_k  
\nonumber \ \  ,\\
A_{[A],k}^{[\dot A]}&
    =&{\rm{tr}}\left (A_{[A]}^{[\dot A]}\lambda_k\right )
\ \  .    
\label{potentialcomponents}\end{eqnarray}
One can then naturally define the generalized covariant derivatives $\cal{D}$
by the matrix
\begin{equation}
{\cal{D}}_{[A]}^{[\dot A]}=X_{[A]}^{[\dot A]}+A_{[A]}^{[\dot A]}
\ \ \ .
\label{covariantderiv}\end{equation}
It is then also natural to define the generalized (in general reducible) 
curvature gauge fields
$\hat F$, as matrices of functions of the coordinates 
(corresponding to the sets $[A],[\dot A]$ and $[B],[\dot B]$),
by the equations
\begin{eqnarray} 
 \hat F_{[A]  [B]}^{[\dot A][\dot B]} 
 =&&\left [{\cal{D}}_{[A]}^{[\dot A]} ~,~
           {\cal{D}}_{[B]}^{[\dot B]}\right ]_{\eta} \nonumber
\\&&  
   -\ \sum_{s=0}^{\min (2a,2b)}\sum_{\dot t=0}^{\min (2\dot a,2\dot b)}
  t(a,\dot a\ ;b,\dot b\ ;a+b-s,\dot a+\dot b-\dot t)\nonumber
\\&& \qquad\qquad \qquad \qquad \quad
 S[A]S[\dot A]S[B]S[\dot B]\epsilon_{[s]}\epsilon^{[\dot t]}
     {\cal{D}}_{[C(s)]}^{[\dot C(\dot t)]}
\ \ \ . 
\label{gaugefield}  \end{eqnarray} 
On the right hand side, the second term has been subtracted 
so as to yield gauge fields free of differential
operators in a gauge-covariant manner. The gauge fields $\hat F$ 
take values in the algebra of the underlying Lie group.

The behaviour of the gauge fields is not irreducible under the
action of the Lorentz group. They decompose according to
\begin{equation} 
\hat F_{[A]  [B]}^{[\dot A][\dot B]}= 
\sum_{s=0}^{\min (2a,2b)}\sum_{\dot t=0}^{\min (2\dot a,2\dot b)}
      S[A]S[B]S[\dot A]S[\dot B]
	       \epsilon_{[s]}\epsilon^{[\dot t]}
		     F_{[C(s)]}^{[\dot C(\dot t)]}
\ \ \ .
\end{equation}
The irreducible components $F_{[C(s)]}^{[\dot C(\dot t)]}$
transforming according to the $(a+b-s , \dot a+\dot b-\dot t)$
Lorentz representations may be projected out by contracting
the curvatures $\hat F$ with the inverses of the generalized 
epsilon tensors $\epsilon_{[s]}$ and $\epsilon^{[\dot t]}$ in
(\ref{genep}),
\begin{eqnarray}
\epsilon^{[s]}&=&
       \epsilon^{\beta_{1}\alpha_{1}}
       \epsilon^{\beta_{2}\alpha_{2}}
       \ldots
       \epsilon^{\beta_{s}\alpha_{s}}
\nonumber \\
\epsilon_{[\dot t]}&=&
         \epsilon^{\dot\beta_1\dot\alpha_1}
         \epsilon^{\dot\beta_2\dot\alpha_2}
         \ldots
         \epsilon^{\dot\beta_{\dot t}\dot\alpha_{\dot t}} 
\\  \epsilon^{[s]} \epsilon_{[s]}&=&  2^s
\ \  \nonumber
\label{epinv}
\end{eqnarray}
and symmetrizing over the remaining multi-indices 
$[C(s)]$ and $[\dot C(\dot t)]$ in (\ref{Csets}): 
\begin{equation} 
  F_{[C(s)]}^{[\dot C(\dot t)]} =  \kappa(s,\dot t)
  S[C(s)]S[\dot C(\dot t)] \epsilon^{[s]}\epsilon_{[\dot t]} 
      \hat F_{[A] [B]}^{[\dot A ][\dot B]}
\ \  ,\label{irgaugefields}\end{equation}
where $\kappa(s,\dot t)$ are combinatorial factors.
The gauge algebra components of the irreducible fields $F$ may be
extracted by taking the traces as in (\ref{potentialcomponents}).  

\section{Generalized self-duality}

Once a model has been defined by the choice of
\begin{itemize}
\item a Lie algebra,
\item a coherent set of coordinates $Y$
and of vector fields $X$ (i.e. having chosen the $t$, $u$ and $c$ 
parameters satisfying the super Jacobi identities),
\end{itemize}
we define a set of generalized self-duality equations as the
vanishing of a subset of the irreducible components
(\ref{irgaugefields}) of the gauge fields.

When $Y$ and $X$ are reduced to belong to the 
$({1\over 2},{1\over2})$ representation of the Lorentz group only 
and the $X_{\alpha\dot\alpha}$ are simply the (commuting among 
themselves) derivatives with respect to the usual Minkowski coordinates
$Y^{\alpha\dot\alpha}$, the irreducible components of the field $\hat F$
defined above are $F_{\alpha\beta}$ and $F_{\dot\alpha\dot\beta}$.
They transform respectively as the (1,0) and (0,1) representations of
the Lorentz group. The usual self-duality equation is then equivalent
to the vanishing of the (0,1) components,
\begin{equation}
F^{\dot\alpha\dot\beta}=0 \ \ \ ,
\label{selfduality}\end{equation}
with the self-dual (1,0) components remaining (and vice-versa for 
anti-self-duality). 

The simplest extension to our generalized superspaces is the vanishing
of all curvature representations except for those transforming
according to the $(a,0)$ representations, 
$\{ F_{\alpha_1\ldots\alpha_{2a}}\}$, containing no dotted indices.
In other words, the conditions
\begin{equation}  \{
F^{\dot\alpha_1\ldots\dot\alpha_{2\dot b}}_{\alpha_1\ldots\alpha_{2a}} 
= 0\  ; 1\le 2\dot b \le \dot t \ \  ,\  0\le 2a\le s   \} ,
\label{hierarchy}\end{equation} 
for any fixed choice of $s,\dot t$.
These systems are solvable in the sense that the traditional 
self-duality conditions (\ref{selfduality}) are.
They provide an infinite hierarchy of self-dual systems incorporating,
for the choice of $X$'s and $Y$'s corresponding to standard superspace, 
the supersymmetric self-duality conditions (see for example \cite{DO}).
 
\section{Conclusion and outlook}
   
By successively defining supercoordinates, a superalgebra of
superderivatives acting linearly on them and finally superpotentials 
in a manifestly Lorentz covariant manner, we have given a natural 
definition of super gauge fields on our generalized superspaces. 
The vanishing of well-chosen irreducible components of the curvature 
superfields is then a natural generalization of self-duality or 
antiself-duality. 
The procedure outlined here will be given in more detail in a
forthcoming paper \cite{DN} where moreover the connections with other
existing generalizations will be made clearer and where further
explicit examples will be given. 

There are many obvious generalizations of this approach. Let us
mention two. One could, for example, try to quantum--deform the
theory by allowing the coordinates to fulfil the commutation
relations of a quantum space, the vector fields to satisfy a quantum
algebra and act on the coordinates in a quantum way, with the
underlying gauge algebra being also q-deformed. Another
generalisation is to alter the Lorentz metric and work with spinors
in spaces (euclidean or otherwise) of arbitrary dimension. 

One of the authors (JN) would like to thank S. Randjbar-Daemi and 
F. Hussain for inviting him to the I.C.T.P. in Trieste and the Fonds
National de la Recherche Scientifique (Belgium) for partial support.
The other (CD) thanks E.S. Fradkin and M.A. Vasiliev for encouraging
discussions.

\end{document}